\documentstyle[aps,preprint]{revtex}
\input psfig
\def\k{\mbox{\bf k}}

\begin{document}
\draft
\preprint{PURD-TH-96-02, OSU-TA-08/96, hep-ph/9603378}
\date{21 March 1996}
\title{Classical decay of inflaton}

\author{S. Yu. Khlebnikov$^1$ and I. I. Tkachev$^{2,3}$}
\address{
${}^1$Department of Physics, Purdue University, West Lafayette, IN 47907 \\
${}^2$Department of Physics, The Ohio State University, Columbus, OH 43210\\ 
${}^3$Institute for Nuclear Research of the Academy of Sciences of Russia
\\Moscow 117312, Russia}

\maketitle

\begin{abstract}
We present the first fully non-linear calculation of inflaton decay. We map 
inflaton decay onto an equivalent classical problem and solve the latter 
numerically. In the $\lambda\phi^4$ model, we find that parametric resonance 
develops slower and ends at smaller values of fluctuating fields, as compared 
to estimates existing in literature. We also observe a number of 
qualitatively new phenomena, including a stage of semiclassical 
thermalization, during which the decay of inflaton is essentially as 
effective as during the resonance stage.
\end{abstract}

\pacs{PACS numbers: 98.80.Cq, 14.80.Mz, 05.70.Fh}

\narrowtext

Amplification of quantum fluctuations and transition to semiclassical
behavior are familiar subjects in inflationary cosmology (for a review of 
the latter, see Ref. \cite{Linde}).
Semiclassical fluctuations produced during inflation can explain the observed
large scale structure of the Universe \cite{structure}. The theory of
their production is by now well developed; for a recent rather general
presentation, see Ref. \cite{PS}. After inflation ends, the scalar field
(inflaton) that was driving it begins to oscillate. 
These oscillations are thought
to lead to particle production, inflaton decay, and eventually to 
reheating of the Universe.

It was only recently realized that phenomena related to large occupation
numbers of Bose particles produced by inflaton decay, such as parametric
resonance, can be significant in the reheating process
\cite{KLS,STB,Betal}. That means that semiclassical phenomena may play as 
important role in reheating as they do in the structure formation.
Of course, the eventual thermalized state will include modes with small
occupation numbers, which are essentially quantal, but initial
stages of the reheating process should admit a semiclassical description.

However, semiclassical description of fluctuations produced by inflaton
decay has not been obtained.
The existing treatments of parametric amplification \cite{KLS,STB,Betal}
employ the standard methods of analyzing particle production, based on
Bogoliubov transformation. These methods allow one to take into account {\em
some} of the back reaction effects of produced particles on the oscillating
zero-momentum ($\k=0$) mode, such as time-dependence of the frequency of 
oscillations, but cannot reliably take into 
account many other non-linear effects that become important certainly no 
later than the frequency change. These effects include scattering of produced 
particles off the $\k=0$ mode, which knocks particles out of the $\k=0$
state; rescattering of the produced particles, which populates
non-resonant modes as well as returns particles back to $\k=0$ 
(Bose condensation) \cite{BC}, etc. 
We generically refer to these effects as {\em rescattering};
we will see that they begin as essentially classical interactions.

Order of magnitude estimates of the amplitude squared
of produced fluctuations $(\delta\phi)^2$ at the end of the resonance stage
were obtained in Refs. \cite{KLS,STB,Betal}. 
It is important to know $(\delta\phi)^2$ more precisely,
because in realistic inflationary models, its magnitude
determines whether any symmetries are restored in non-thermal regime after
the resonance stage \cite{pt}. 
This, in turn, requires taking into account all rescattering
effects. After the parametric resonance ends, rescattering leads to chaotic 
behavior and begins to thermalize the system, while it is still in 
the semiclassical regime. This semiclassical thermalization, which should 
proceed rather fast, due to the still large occupation numbers, is very 
important for determining the reheating temperature. 
However, it is completely absent from the standard treatment \cite{KLS,STB},
as well as from the one augmented by time-dependent Hartree approximation
\cite{Betal} (in the Hartree approximation, chaotic behavior characteristic 
of a non-linear classical system is lost).

It is clear, on the other hand, that if one succeeds in finding an adequate
semiclassical description of fluctuations produced by inflaton decay, 
the decay can be studied via the {\em classical} equations of motion and thus
treated as a fully non-linear problem, without any further assumptions. 
That classical problem 
can be turned over to a computer. To obtain the semiclassical description,
one has to find the initial conditions for the classical equations 
of motion, corresponding to the original quantum state. 
We find these initial conditions using the following observation.
Because inflationary models have small coupling constants,
fluctuating modes can grow to large occupation numbers before 
their interactions become important.
In other words, the transition to classical behavior occurs in a linear regime 
with respect to fluctuations.
We can then use the linear theory of quantum-to-classical transitions
\cite{PS} to obtain the initial conditions for subsequent non-linear 
classical evolution.
Possibility of a semiclassical formulation of the problem of inflaton
decay was noted in a recent paper by Son \cite{Son}; however, the correct
initial conditions for the classical problem were not obtained there.

In this Letter we report the first results of our numerical study of
the semiclassical decay of inflaton, for the simplest model of a real scalar
field with the $\lambda \phi^4$ potential. 
Initial conditions were chosen to correspond
to conformal vacuum for fluctuations at the end of inflation \cite{f1}.
Numbers for various quantities of physical interest are presented below; 
because ours is the first fully non-linear
treatment of the problem, they differ from estimates existing 
in the literature. We also observe a number of qualitatively new phenomena,
including the stage of inflaton decay when the resonance peaks in power 
spectrum begin to interact and smear out; we call this stage semiclassical 
thermalization. We expect this stage on general grounds, as noted above,
but it was not observed previously. We find that during semiclassical 
thermalization inflaton decay is essentially as effective as during the
stage of parametric resonance.

We consider a real scalar field minimally coupled to gravity,
in a Friedmann-Robertson-Walker universe.
We work with rescaled variables: time $\tau$, such that 
$a(\tau) d\tau=\sqrt{\lambda}\phi_0(0)a(0)dt$, coordinate 
$\mbox{\boldmath $\xi$}=\sqrt{\lambda} \phi_0(0)a(0) \mbox{\bf x}$, and 
field $\varphi=\phi a(\tau)/\phi_0(0) a(0)$, where $a(\tau)$ is the scale 
factor, and $\phi_0(0)$ is the zero-momentum mode at $t=\tau=0$.
In this Letter we consider only the massless inflaton case. 
(The method itself applies to other models, as well as to various laboratory
systems.) The action is 
\begin{equation}
S= \frac{1}{\lambda}\int d^3\xi d\tau \left[
  {1 \over 2} \left( \dot \varphi -\frac{\dot a}{a} \varphi \right)^2 
                         - \frac{(\nabla_{\xi} \varphi)^2}{2} - 
   \frac{\varphi^4}{4} 
                      \right]
\label{act'}
\end{equation}
where dots denote derivatives with respect to $\tau$. The parameter
${\dot a}/{a}\equiv h$ is the rescaled Hubble parameter,
$h(\tau)=H(\tau)/\sqrt{\lambda}\phi_0(0)$. For the massless inflaton, 
$H(0)\sim \sqrt{\lambda}\phi_0^2(0)/M_{\rm Pl}$, so that $h(0)$ is independent
of $\lambda$ and is of order one.
The equation of motion following from (\ref{act'}) is
\begin{equation}
{\ddot \varphi} - \nabla_{\xi}^2 \varphi -\frac{\ddot a}{a} \varphi + 
\varphi^3 = 0 \; .
\label{equm}
\end{equation}
When the massless field begins to oscillate, the Universe becomes radiation 
dominated, so that ${\ddot a}=0$, and the corresponding term drops out 
of Eq. (\ref{equm}). 
Although the coupling $\lambda$ does not appear in (\ref{equm}), it
appears in the action and hence regulates the magnitude of initial
quantum fluctuations in non-zero modes relative to the magnitude of the 
zero mode. We have checked numerically that in the massless case 
the condition for a semiclassical regime to occur is precisely the smallness 
of the coupling $\lambda$. The zero mode is semiclassical from the beginning.
Its initial value is $\varphi_0(0)=1$, by virtue of our rescaling. 
It is possible to choose ${\dot \varphi_0}(0)=0$ as a definition of 
the moment of time when the oscillations start; we do that.
We make no further assumptions about the zero mode: its time dependence
will come out of the solution to the full non-linear problem, once we specify 
the initial conditions for non-zero modes.

According to the preceding discussion, to obtain the initial data for the
classical problem, we can linearize Eq. (\ref{equm}) with respect to
fluctuations. The linearized equation of motion for the Fourier transform of
the fluctuation field ($\k\neq 0$) is
\begin{equation}
{\ddot \varphi}_{\k} + \omega_k^2(\tau) \varphi_{\k} = 0
\label{equ}
\end{equation}
where $\omega_k^2(\tau) = 3\varphi_0^2(\tau) + k^2$.
The operator solution to Eq. (\ref{equ}) is $\varphi_{\k}(\tau) = 
f_{k}(\tau) b_{\k}(0) + f^*_{k}(\tau) b^{\dagger}_{-\k}(0)$,
where $c$-number function $f_{k}(\tau)$ satisfies Eq. (\ref{equ})
and the initial conditions
\begin{equation}
f_{k}(0)=\left( \frac{\lambda}{2\omega_k(0)} \right)^{\frac{1}{2}} ;
{}~~~{\dot f}_{k}(0)=
\left[ -i\omega_k(0) + h(0) \right] f_k(0) \; .
\label{ini}
\end{equation}
Operators $b^{\dagger}$ and $b$ are the creation and annihilation 
operators defined at zero time and normalized according to 
$[b_{\k},b^{\dagger}_{\k'}]=\delta(\k-\k')$. 
A mode becomes semiclassical if the corresponding function
$f_k(\tau)$ grows (exponentially) for large $\tau$ (and 
so does, then, the corresponding occupation number).  
In that case, it can be shown \cite{PS} that up to exponentially small
corrections, $f_k$ at large $\tau$ can be made real by a time-independent
phase rotation. When the exponentially small terms are neglected,
$\varphi_{\k}(\tau)$ and its canonical momentum commute,
and therefore $\varphi_{\k}(\tau)$ at large times can be regarded 
as a random classical variable. This is the semiclassical 
description.

The distribution of values of $\varphi_{\k}(\tau)$ is obtained from 
the solution to the linearized quantum problem. It depends on the quantum 
state for fluctuations existing at the end of inflation. 
Modes with $k\sim 1$ (in our rescaled units), which we are 
interested in, could only be produced towards the end of the de Sitter epoch, 
so the quantum state for these modes should not be much different from 
conformal vacuum. Below we present results obtained using conformal vacuum
as the original quantum state \cite{f1}. The corresponding classical 
distribution function at large times is, cf. Ref. \cite{PS},
\begin{equation}
{\cal F}[\varphi, {\dot \varphi};\tau] = {\cal N} 
  \exp\left(- {\int}' \frac{|\varphi_{\k}|^2}{|f_k(\tau)|^2} d^3k \right)
\delta
  \left(f_k(\tau) {\dot \varphi}_{\k}- {\dot f}_k(\tau) \varphi_{\k} \right)
\label{dis}
\end{equation}
where the delta function is a shorthand for the product of functional
delta functions for real and imaginary parts, and the prime on the integral 
reminds us that 
the integral is taken over half of the values of $\k$ (since these give 
all independent $\varphi_{\k}$) and does not include $\k=0$;
${\cal N}$ is a time-independent normalization.
A useful point here is that (\ref{dis})
satisfies the classical Liouville equation also at small times. So, we can 
use it as an initial condition even at $\tau=0$, where $f_k$ and its 
derivative are given by Eq. (\ref{ini}). 

We thus use a discretized version of the distribution
\begin{equation}
{\cal P}[\varphi(0)] = N \exp\left( -\frac{2}{\lambda} {\int}' \omega_k(0)
            |\varphi_{\k}(0)|^2 d^3 k \right)
\label{dis'}
\end{equation}
to generate random initial values
of $\varphi_{\k}$. Once a value of $\varphi_{\k}$ is generated, 
the corresponding
"velocity" is determined uniquely, according to eqs.(\ref{dis}) and
(\ref{ini}), 
as ${\dot \varphi}_{\k}(0) = \left[-i\omega_k(0) + h(0) \right]
\varphi_{\k}(0)$. 
Together with the initial conditions for the zero mode,
this forms (after an inverse Fourier transform) an initial data problem for 
the fully non-linear equation (\ref{equm}).

We have integrated numerically Eq. (\ref{equm}) in a box of finite size $L$
with the above initial conditions. We have varied the 
size of the box and the number of grid points $N$ to make sure that we 
are close to the continuum limit. 
We have varied the coupling constant $\lambda$
over the range $10^{-10}$--$1$. Here we present the results
for $L=16 \pi$, $N=128^3$ and $\lambda =10^{-4}$.
All quantities we measured are averages of some sort; for these quantities,	
there is no need in further averaging over initial conditions. 

\begin{figure}
\psfig{file=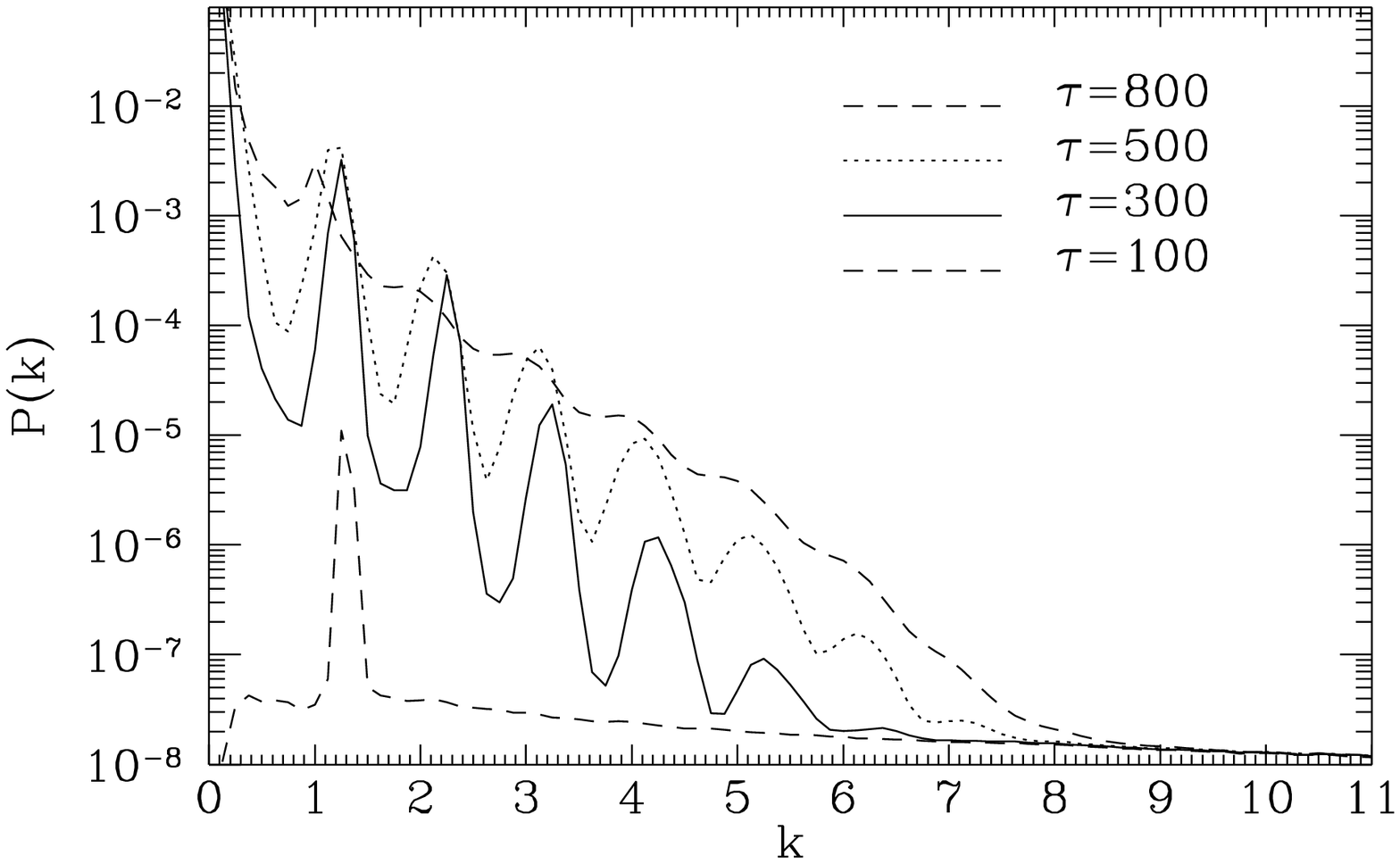,height=3.7in,width=6in}
\caption{Power spectrum of fluctuations at successive moments of time.}
\label{fig:pws}
\end{figure}

One important quantity that can be measured is the power spectrum
of fluctuations, $P(k)$. It is proportional to
$\varphi^*_{\k} \varphi_{\k}$ averaged over the direction of $\k$ and
is normalized in such a way that Parseval's theorem reads
\begin{equation}
\int d^3 k P(k) = 
{1 \over L^3} \int d^3 \xi [\varphi(\xi)-\varphi_0]^2 \; .
\label{par}
\end{equation}
Thus normalized power spectrum does not depend on the size of the 
(large) box. 
It is presented in Fig. \ref{fig:pws} for several moments of time.
Resonance peaks develop at $\tau< 200$ in the following order: 
first at $k \approx 1$, 
somewhat later at $k$ close to zero, and still later at larger momenta.
For comparison, we have run the problem linearized with respect to
fluctuations. Only the initial development of the first peak was unaffected 
by linearization. The peak at $k \approx 2$ was barely visible, the others
were not seen at all. We conclude that all peaks in Fig. \ref{fig:pws}
except the one at $k \approx 1$ are due to rescattering. 
The peak at $k$ close to zero is not expected from the approximate
linear analysis based on the Mathieu equation. 
We interpret it as a result of the rescattering process in which a particle 
from the first peak transfers some 
of its momentum to a particle from the $\k=0$ condensate \cite{f2}.
The positions of the $k\approx 2$ and higher peaks are close to those 
predicted by the Mathieu equation, but their widths and magnitudes are not. 
We interpret them as another result of rescattering: for example, two 
particles from the first peak, with $k\approx 1$, together with two particles 
from the condensate produce two particles with $k\approx 2$. (That process
may be more efficient than the $6\to 2$ process, which uses only condensate
particles and is approximately described by the Mathieu equation.) 
Thus, rescattering processes start to play some role already at 
the parametric resonance stage. 

\begin{figure}
\psfig{file=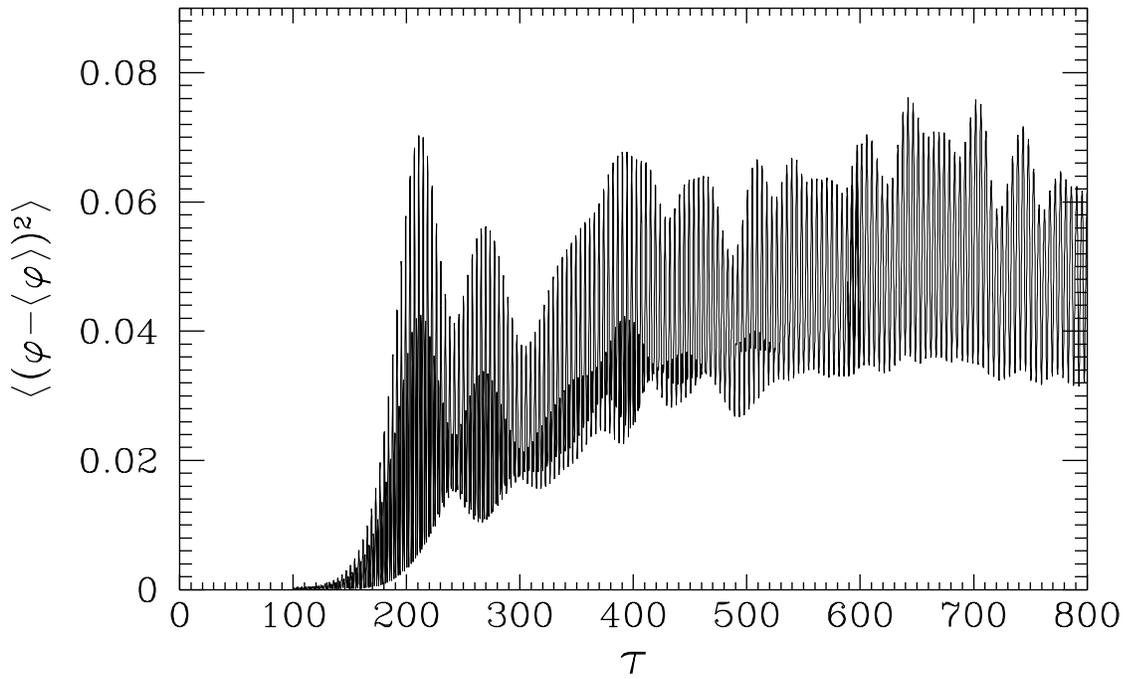,height=3.8in,width=6in}
\caption{Variance of the scalar field as a function of time.}
\label{fig:phi2}
\end{figure}

Another interesting quantity is the integral of the power spectrum of produced 
fluctuations over all (non-zero) $k$. This is simply the variance
$\mbox{Var}(\varphi)= \langle (\varphi(x)-\langle\varphi\rangle )^2\rangle$ 
and can be measured independently. The brackets denote averaging over 
the spatial lattice, so $\langle \phi \rangle$ is just the zero-momentum mode. 
The lattice effectively subtracts contribution of high-momentum modes, 
so at small $\lambda$ our $\mbox{Var}(\varphi)$ approximates an already 
subtracted continuum
quantity, equal to the continuum $\mbox{Var}(\varphi)$ minus its value at
zero time. The behavior of
$\mbox{Var}(\varphi)$ with time is shown in Fig. \ref{fig:phi2}.
Initially it grows exponentially; this is the stage of parametric resonance.
Using logarithmic plots, we have found the following analytical fit for this
stage
\begin{equation}
\ln\mbox{Var}(\varphi)=2\mu\tau + \ln\lambda - 8.
\label{fit}
\end{equation}
Because $\mbox{Var}(\varphi)$ oscillates, the choice of constant here is
somewhat arbitrary; we chose it in such a way that the line (\ref{fit})
passes through the maximum of the amplitude of $\mbox{Var}(\varphi)$ 
at the end of the resonance stage
(located at $\tau\equiv\tau_* \approx 200$ if $\lambda =10^{-4}$). 
For the rate of the exponential growth $2\mu$ we obtain $2\mu=0.07$, for all
small enough $\lambda$. This value is smaller than the one obtained  
from the Mathieu equation arising when the time dependence 
of the zero mode (an elliptic function in this regime) is replaced by sine
with the same period.
When we make this replacement in our program, we obtain for the rate the value
of 0.11, in our time units, as expected. As another test, we have varied the size of the
box and verified that we do not miss the fastest growing mode.

The value of $\mbox{Var}(\varphi)$ in the maximum
at the end of the resonance stage 
is approximately $0.07$ and does not depend on $\lambda$ when $\lambda \ll 1$.
Thus, for general (small) $\lambda$ the resonance ends at time 
$\tau_* = 76. - 14.3 \ln\lambda$.
Extrapolating to a realistic value $\lambda =10^{-13}$, 
we find $\tau_*\approx 500$. 
To obtain the variance of the physical field $\phi$ at time $\tau$, we need, 
at large $\tau$, to multiply
$\mbox{Var}(\varphi)$ by $3M_{\rm Pl}^2/2\pi\tau^2$. Hence, the maximum
variance of $\phi$, at the end of the resonance stage, is 
\begin{equation}
\mbox{Var}(\phi) = 1.\times 10^{-7} M_{\rm Pl}^2
\label{phy}
\end{equation}
For the effective temperature $T_{\rm eff}$, defined by equating
(\ref{phy}) to $T_{\rm eff}^2/12$, this gives $T_{\rm eff}= 1.\times 10^{-3}
M_{\rm Pl}$. 
Notice that $T_{\rm eff}$ does not depend on the initial value of the
inflaton field after inflation.
The smallness of $T_{\rm eff}$ is a result of two factors 
not taken into account previously: 
$\mbox{Var}(\varphi)\ll 1$ and very large $\tau_*$.
In realistic models including more fields, $T_{\rm eff}$
determines if any symmetries are restored in the non-thermal regime
after the resonance stage \cite{pt}. If in those models
$T_{\rm eff}$ remains three orders of magnitude lower than the Planck scale, 
then, for example, non-thermal restoration of GUT symmetry may be prevented.
On the other hand, we see that $T_{\rm eff}$ is much larger than the reheating
temperature that would be obtained if one neglected the effect of large 
occupation numbers, which confirms the main point of Refs.\cite{pt}.

\begin{figure}
\psfig{file=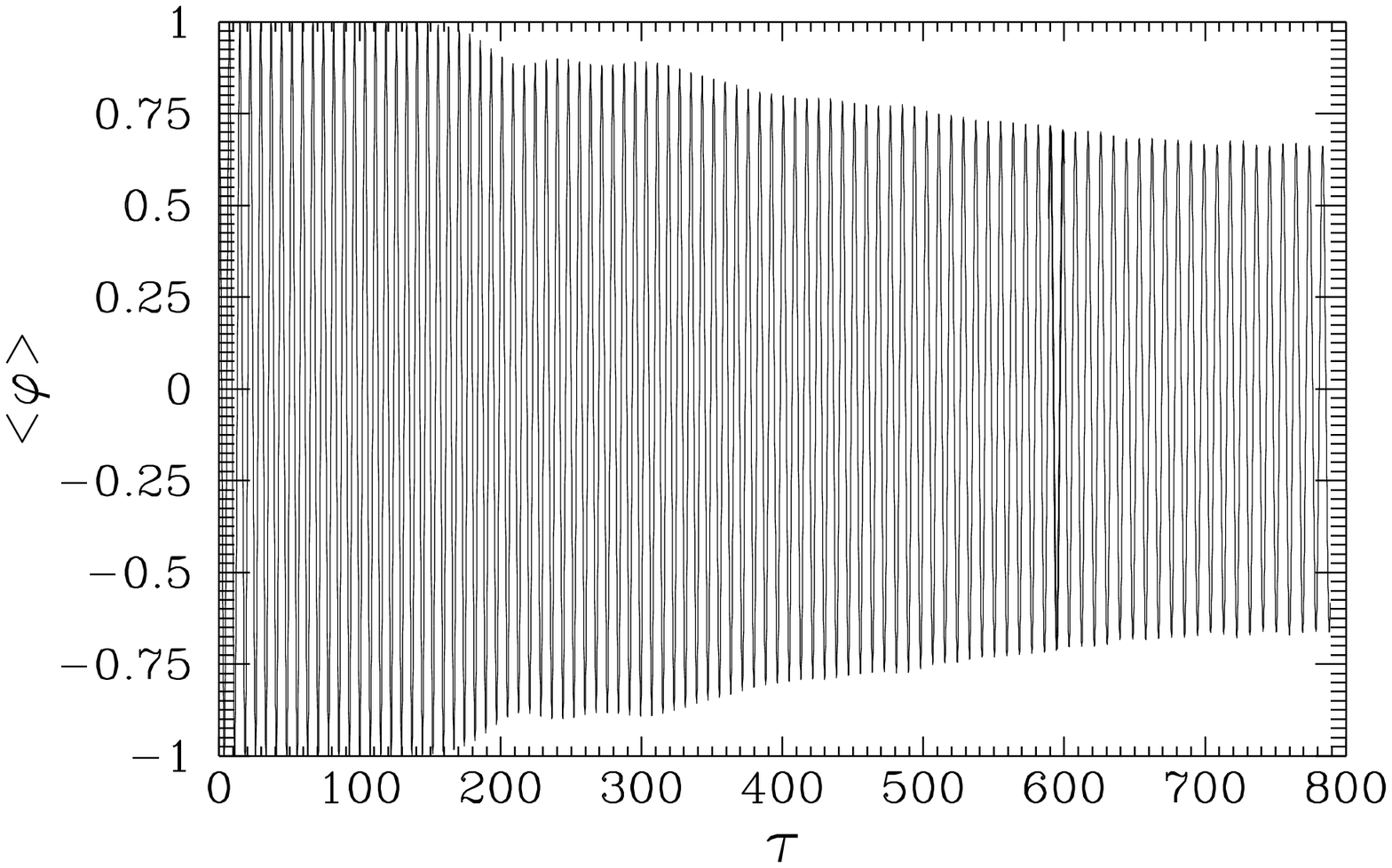,height=3.8in,width=6in}
\caption{Time dependence of the zero-momentum mode.}
\label{fig:zmod}
\end{figure}

Finally, the time dependence of the zero-momentum mode is shown in Fig. 
\ref{fig:zmod}. We can see that after decreasing during 
the parametric resonance, it rebounces twice during time $\tau=200$ to 300. 
We attribute this to Bose condensation caused by rescattering. 
Comparison of Figs. \ref{fig:phi2} and \ref{fig:zmod} suggests that it
is Bose-condensation that terminates parametric resonance.
However, the magnitude
of the zero mode continues to decrease significantly, approximately
as $\tau^{-1/3}$ (as opposed to a slower $\tau^{-1/6}\propto t^{-1/12}$ decay 
found in ref. \cite{KLS}). Returning to Fig.
\ref{fig:pws} we see that the peaks smear out at this stage.
This is the onset of classical chaos or, in the language of particle physics, 
stimulated scattering, decay and annihilation due to large 
occupation numbers.
This process is rather effective. In our model the energy scales as the forth
power of the field. So, for $\lambda=10^{-4}$, at time $\tau \approx 400$ 
about 70\% of energy is in fluctuations.
For smaller $\lambda$, the onset of this chaotic regime, which we call
semiclassical thermalization, is delayed, but its rate with respect to
time $\tau$ does not 
decrease. The reason is that the strength of rescattering is determined not
by $\lambda$ itself but by
the product of $\lambda$ and a relevant occupation number.

In conclusion, we have done the first fully non-linear calculation of inflaton 
decay. We have mapped inflaton decay onto an equivalent classical problem
and solved the classical problem numerically. In the $\lambda\phi^4$ model,
we have found that parametric resonance
develops slower and ends at smaller values of fluctuating fields,
as compared to estimates existing in the literature. We have also observed 
a number of qualitatively new phenomena, including a stage of semiclassical
thermalization (chaos), during which the decay of inflaton is essentially as
effective as during the resonance stage.

I.T. thanks A. Riotto for discussions. The work of S.K. was supported
in part by the U.S. Department of Energy under grant DE-FG02-91ER40681 
(Task B), by the National Science Foundation under grant PHY 95-01458, 
and by the Alfred P. Sloan Foundation. The work of I.T. was supported
by DOE grant DE-AC02-76ER01545 at Ohio State.

\begin{thebibliography}{99}
\bibitem{f1}
We tried other distributions as well; the results are rather
insensitive to the form of initial distribution, as long as 
the initial fluctuations are of correct order of magnitude.
\bibitem{f2}
This is apparently the strongest rescattering process at this stage, because 
it takes advantage both of the existence of the condensate and of the already
large occupation number in the first peak.
\end{thebibliography}
\end{document}